\documentstyle[floats,pra,aps]{revtex}
\tightenlines 
\begin{document} 

\title{\large\bf The World of Synchrotrons\footnote{\normalsize
In {\em Resonance}, {\bf 6}, No. 11, pp. 77--86 (November 2001). \\
A monthly Publishon of the {\bf Indian Academy of Sciences}, 
http://www.ias.ac.in/resonance/
}}

\author{Sameen Ahmed KHAN}
\address
{
12-2-/830/25/2, Alapati Nagar, Mehdipatnam, Hyderabad 500028 \\
khan@pd.infn.it ~~~ http://www.imsc.ernet.in/$\sim$jagan/khan-cv.html \\
}
\maketitle

\medskip

\noindent
{\bf Abstract:}
A summary of results on synchrotron radiation is presented along with 
notes on its properties and applications.  Quantum aspects are briefly
mentioned.  Synchrotron radiation facilities are described briefly 
with a detailed coverage to the accelerator programmes in India.  The
relocated and other upcoming synchrotrons are also described in some
detail.

\medskip

\section{Introduction to Synchrotron Radiation}
Charged-particles when accelerated radiate electromagnetic energy.  
This interesting physical phenomenon, now known by the name 
{\em synchrotron radiation} had its theoretical beginnings a long
time ago at the time of classical electrodynamics.  At that time,
only the very basic features of this physical phenomenon were studied
and expressions were derived for several quantities such as the
{\em total radiation intensity}, {\em spectral distribution} and
{\em angular distribution}~\cite{Jackson}.  These theoretical studies 
had to wait for about half a century till the development of 
charged-particle accelerator technology for a direct observation and 
experimental verification.  It was experimentally observed for the
first time in 1947 in the $70 MeV$ electron synchrotron and hence the
name {\em synchrotron radiation}.  This observation generated a renewed
interest in synchrotron theory.  Synchrotron radiation was an
irritant in early electron synchrotrons and storage rings.  But it 
was soon realized that synchrotron radiation was a very valuable
product in itself for research applications requiring intense and
bright sources of light over a wide range of wavelengths.  Electrons
loose a large amount of energy in the form of synchrotron radiation
putting a limit on the maximum attainable energy in a given type of 
accelerator.  Let us first consider the case of the betatron.  The
betatron is a cyclic electron accelerator with a circular orbit of
approximately constant radius which provides acceleration through
{\em magnetic induction}.  As the beam energy rises synchrotron
radiation losses rise and begin to compete with the energy gained due
to magnetic induction.  In practice, the synchrotron radiation begins
to become important at about $100 MeV$ and limits beam energies to
about $300 MeV$.  This challenge of the synchrotron radiation
stimulated the development of accelerator technology and further
increased the energy of the accelerated particles.

\medskip

Here, it would be relevant and interesting to mention the very
{\em exceptional} case of the charged-particles under uniform
acceleration, {\em i.e.,} a constant force.  This constant force
can be produced, for example, by a constant uniform electrostatic 
field.  From the special theory of relativity we know that a particle
under a constant force executes hyperbolic motion.  Does a
charged-particle under uniform acceleration (hyperbolic motion) 
radiate?  The answer to this question is not yet completely resolved!
This topic is listed as one of the several
{\em surprises in theoretical physics} in the compilation due to
Peierls~\cite{Peierls}.

\medskip

One of the characteristics of the synchrotron radiation is its
intensity (energy emitted per unit time).  Let us first consider
the motion of a charged-particle of rest mass $m_0$ and charge $q$
in a uniform magnetic field of strength $B$ with the simplification
that there is no component of the velocity along the field 
direction.  We further decide to neglect the changes in the
trajectory due to the radiation losses.  In such a configuration the
particle moves in a circle.  The radius $R$ of this circle is given
by the well-known relation
\begin{equation}
R 
= \beta \gamma \frac{m_0 c}{q B} 
= \frac{\beta}{c} \frac{E}{q B}
\end{equation}
where the total energy $E = \gamma m_0 c^2$, $c$ is the velocity of 
light, $\beta = v/c$ and the relativistic factor
$\gamma = \frac{1}{\sqrt{1 - \beta^2}}$.  In 1898 Li\'{e}nard derived
the expression for the total radiation intensity $P$ for a 
charged-particle moving in circular motion~\cite{Chao}
\begin{equation}
P
= \frac{2}{3}
\frac{q^2 c}{4 \pi \epsilon_0 R^2} \beta^4 
\left(\frac{E}{m_0 c^2}\right)^4
\label{Power}
\end{equation}
The radiated power depends on the rest mass of the radiating particle 
like $1/{m_0^4}$.  For protons and electrons of the same total energy
$E$ the ratio of the radiated powers is
\begin{equation}
\frac{P_p}{P_e}
=
\left(\frac{m_e}{m_p}\right)^4 
=
8.80 \times 10^{- 14}
\end{equation}
Synchrotron radiation is the dominant factor in the design of high
energy electron synchrotrons and is an obstacle to exceeding
$100 GeV$ or so in this type of accelerator.  Only now synchrotron
radiation is becoming a design consideration for proton synchrotrons.
In the proton case, single-particle motion, to a very good
approximation, exemplifies a Hamiltonian system.  Particle motion in
electron synchrotrons, on the other hand, is inherently dissipative.

\medskip

\section{Quantum Effects in Synchrotron Radiation}
Synchrotron radiation was experimentally observed in a period when 
there was a very keen interest in analyzing the quantum corrections to
the prescriptions based on the classical theories.  The need of such 
studies also arose from the desire to achieve higher beam energies with
the evolving accelerator technology.  A quantum mechanical expression
for the radiation intensity was derived by Schwinger\cite{Schwinger}
which is:
\begin{equation}
P^{\rm Quantum}
=
P^{\rm Classical}
\left(1 - \frac{55}{8 \sqrt{3}} \frac{\epsilon}{E} + \cdots
\right)
\end{equation}
where the critical energy of the radiated photon
$\epsilon = \hbar \omega_c 
= \frac{3}{2} \gamma^3 \frac{\hbar c}{R}$.  For an electron the
quantum contributions become effective only at about $10^{4} GeV$.
There have been also studies to assess the effect of spin and the
anomalous magnetic moment on the radiation intensity.  These effects
are also noticeable only at very high energies.  As far as the
radiation intensity is concerned the quantum contributions are of
no consequence in any realizable accelerator.  However, there are
several other ways by which quantum effects manifest at energies
realizable in many of the accelerators.  The {\em quantum radiation
fluctuations} start having an appreciable effect on the motion of
the particles at energies exceeding
$E_c = m_0 c^2 \left(\frac{m_0 c R}{\hbar}\right)^{1/5}$
which for an electron is about $500 MeV$~\cite{Ternov}.  Quantum
fluctuations of the radiation have to be taken into account in the
engineering calculations of the particle motion.  In passing it is to
be noted that the quantum corrections to the {\em beam optics} are
related to the powers of the de~Broglie wavelength of the 
charged-particle.  Hence the quantum corrections to the beam optics
are more noticeable at lower energies~\cite{Jagannathan}.  One 
practical aplication of the {\em quantum formalism of charged-particle
beam optics} would be to get a deeper understanding of the polarized
beams.

\medskip

It has been confirmed that the synchrotron radiation is responsible 
for the directional orientation of the particle spin {\em i.e.,} it
leads to {\em radiation self-polarization} of the beam.  As a result
of the quantum fluctuations of the synchrotron radiation, the 
particle spin achieves a state whose orientation direction is 
opposite to that of the magnetic field.  The particle beam becomes
$92\%$ polarized in a time $\tau$ given by
\begin{equation}
\frac{1}{\tau}
=
\frac{5 \sqrt{3}}{8}
\frac{\hbar}{m_0 c R} \gamma^5
\frac{q^2}{4 \pi \epsilon_0 m_0 c R^2}\,.
\end{equation}
For many realizable accelerators this time turns out to be about few 
hours and hence it is possible to observe (and utilize) radiation
self-polarization in numerous storage rings.  The radiation 
self-polarization is currently the only method of obtaining 
relativistic beams with an oriented spin.  A novel proposal to produce
polarized beams using the proposed spin-splitter devices based on the
Stern-Gerlach kicks has been presented recently~\cite{Pusterla}.  
Polarized beams are very essential for many of the experiments in
particle physics.

\medskip

Some of the many important properties of the synchrotron radiation
are summarized below:

\begin{enumerate}

\item
The angular distribution of the synchrotron radiation is very
sharply peaked in the direction of the particle's velocity vector
within an angular width of $1/{\gamma}$.  The radiation is
plane-polarized on the plane of the particle's orbit, and 
elliptically-polarized outside this plane.

\item
The radiation spans a continuous spectrum.  The power spectrum
produced by a high energy particle extends to a critical frequency
$\omega_c = \frac{3}{2} \gamma^3 \omega_R$ where the cyclotron
frequency $\omega_R = \frac{c}{R} = \frac{q B}{\gamma m_0}$ 

\end{enumerate}

\medskip

These results imply that the synchrotron radiation is extremely 
intense over a broad range of wavelengths from the infrared through 
the visible and ultraviolet range and into the vacuum ultraviolet and 
soft and hard X-rays parts of the electromagnetic spectrum.  The high
intensity over a very broad spectrum range and certain other 
properties (including, collimation, polarization, pulsed-time 
structure, partial coherence, high-vacuum environment, ...) make
synchrotron radiation a very powerful tool for a variety of
applications in basic and applied research and technology.  It is
particularly important in those parts of the electromagnetic spectrum
where laser sources are (presently) not available such as the vacuum 
ultraviolet, soft and hard X-rays, parts of the infrared, {\em etc.,}. 
The applications of the synchrotron light span a wide range of domains
in fundamental science (chemistry, physics, biology, molecular
medicine, {\em etc.,}) applied research (materials science, medical 
imaging, pharmaceutical R \& D, advanced radiology, {\em etc.,}) and 
industrial technology (micro-fabrication, micro-analysis, 
photo-chemistry, {\em etc.,}). 

\medskip

\section{Numerical Estimates}
Using Equation~(\ref{Power}) let us estimate the energy radiated by a
single particle in one revolution.  The time, $T$ of one revolution is
${2 \pi R}/{\beta c}$ and the energy ${\cal E}$ lost is
\begin{eqnarray}
{\cal E} 
& = & P \times T \nonumber \\
& = &
\frac{1}{3} \frac{q^2}{\epsilon_0 R} \beta^3 \gamma^4\,.
\end{eqnarray}
The maximum energy a particle can loose by radiation is all its
kinetic energy, $K = (\gamma - 1) m_0 c^2$.  This can happen only at
ultrarelativistic energies ($\gamma \gg 1$ and $\beta \approx 1$).  
The required $\gamma$ is found by equating ${\cal E}$ to $K$.  Then
one gets
\begin{equation}
\gamma
=
\frac{1}{\beta}
\left[
3 m_0 c^2 \frac{\epsilon_0 R}{q^2} \right]^{\frac{1}{3}}\,.
\end{equation}
In the particular cases of an electron and a proton one gets
\begin{eqnarray}
K_e & = & 20 R^{\frac{1}{3}} GeV\,, \nonumber \\
K_p & = & 5 \times 10^{5} R^{\frac{1}{3}} GeV\,, 
\end{eqnarray}
where $R$ is in $meters$.  In any realizable accelerator $R$ is about
several $km$ which limits the energy to several $100$'s $GeV$ for an
electron-synchrotron and to about a thousand $TeV$ for a
proton-synchrotron.  So one needs to explore other types of machines.
For high-energy $e^{+} e^{-}$ colliders, {\em linear accelerators}
become a very attractive option.  This is why all $p \bar{p}$ colliders
are circular and all future high-energy $e^{+} e^{-}$ colliders will
likely be linear.  In the high-energy machines several $mega watts$ of
power is dissipated in the form of synchrotron radiation around the
ring.  This power loss is comparable to the power requirements of a
small town.

\medskip

\section{Synchrotron Facilities} 
A synchrotron radiation facility is based on the technology of 
charged-particle accelerators.  Bunches of charged-particles
(usually, electrons) are made to circulate for several hours inside a
ring-shaped, long tube under high vacuum.  These rings have several
beam lines with experimental stations and serve several sets of users
simultaneously.  Contrary to the expectation there are not very many
synchrotron facilities to meet the huge demands of numerous users.  
This is due to the high costs (about a hundred million US~\$) and the 
required optimum technological expertise.  Currently, around the world
there are about fifty storage rings in operation as synchrotron
radiation sources, located in twenty-three countries.  About a dozen
are under construction and another dozen or so are being planned.  In
Asia there are about twenty synchrotrons laboratories in nine 
countries:  Armenia, China, India, Japan, Jordan, Korea, Singapore, 
Taiwan and Thailand.  This small list leaves out not only many 
countries but the regions (such as the Continents of Africa and 
Australia) without a single synchrotron facility.  India has the 
expertise and the experience of indigenously building two synchrotrons
at the Centre of Advanced Technology ({\bf CAT}) in Indore.  
{\bf Indus-I} is a $450 MeV$ synchrotron and {\bf Indus-II} is a very
energetic $2.0 GeV$ synchrotron. 

\medskip

Storage rings are very flexible devices.  By reusing most of the major
components their performance can be upgraded at an incremental cost 
that is small as compared with the cost of construction of a new 
synchrotron.  In recent years this flexibility is being innovatively
exploited to relocate the very generously donated synchrotrons to 
those locations which are under-represented in the 
{\em World Synchrotron Map}\cite{Synchrotron-Map}. 

\medskip

\section{The Most Powerful Synchrotrons}
The $8.0 GeV$ {\bf SPring-8} synchrotron is the largest synchrotron
radiation source in the world and was completed in 1997 in Hyogo 
prefecture of Japan.  Japan is one of the leading countries along with
the United States and the European Union in accelerator-based science
research.  The SPring-8 belongs to the category of hard X-rays
machines, along with the $7.0 GeV$ Advanced Photon Source ({\bf APS})
in Argonne, USA and the $6.0 GeV$ European Synchrotron Radiation
Facility ({\bf ESRF}) in Grenoble, France.  Owing to their extremely 
high energy these synchrotrons have their specific problems, and have
forced the development of new techniques and new devices in the field
of optics and detectors to ensure the required high stability of the
electron beam.  In view of the very unique challenges arising due to
the very high energy the three most powerful synchrotron laboratories
have signed a 
{\em Framework of Agreement for Collaboration}~\cite{EPS}.

\medskip

\section{Accelerator Programmes in India}
The accelerator programmes in India have a very long history.  It had
an early beginning in 1940 when Prof. Meghnad Saha developed a 
$37 inch$ cyclotron at the Calcutta based Institute of Nuclear Physics,
which is now called Saha Institute of Nuclear Physics ({\bf SINP}).  
In 1950, a $1.0 MeV$ cyclotron was commissioned at the Tata Institute
of Fundamental Research ({\bf TIFR}) in Mumbai.  In 1960, a $5.5 MeV$
Van-de-Graff accelerator was installed at the Bhabha Atomic Research
Centre ({\bf BARC}) in Mumbai.  In 1978, an indigenously designed and 
built $224 cm$ diameter Variable Energy Cyclotron was made operational
at the Variable Energy Cyclotron Centre ({\bf VECC}) in Calcutta.  Now,
there are several Pelletrons such as the $6.0 MeV$ Pelletron at the 
Institute of Physics ({\bf IOP}) in Bhubaneshwar and the $14.0 MeV$
Pelletron at TIFR.  The Nuclear Science Centre ({\bf NSC}) in New Delhi
has a very energetic $150 MeV$ Pelletron, which can accelerate very
heavy ions to high energies.  Very recently a beam of the radioactive 
isotope beryllium-$7$ was produced at the NSC.  This marks India's 
entry into an elite group of nations which are doing research with
radioactive-ion beams~\cite{Radioactive}.  India has the  expertise
and the experience of indigenously building the two synchrotrons at
the Centre of Advanced Technology ({\bf CAT}) in Indore. {\bf Indus-I}
is a $450 MeV$ machine and {\bf Indus-II} is a very energetic
$2.0 GeV$ machine~\cite{CAT-Website}. 

\medskip

\section{Accelerator Meetings in India}
India, one of the very few countries which very regularly holds 
Accelerator and Beam Physics Meetings.  For several years the Centre 
for Advanced Technology ({\bf CAT}) at Indore has been holding a 
series of Schools on the {\em Physics of Beams} every year in 
December--January.  This series of Schools is funded by the Department 
of Science and Technology ({\bf DST}), with the aim of dissemination
more widely in India, knowledge of, and interest in, beam physics.
The School attracts several speakers from the premiere accelerator
laboratories around the world.  The Schools are very well-structured
with tutorials and a few laboratory experiments.  The participants of
the School are further attracted to the 
{\em Summer Research Programme} conducted at CAT (see the very detailed
School Reports in Ref.~\cite{CAT-Schools}).  For over a decade the
IUC-DAEF Calcutta Centre has been holding the tri-annual
{\em National Seminars on Physics and Technology of Particle
Accelerators and their Applications}.  This series of National Seminars
known by the acronym {\bf PATPAA} provide a forum where all the
accelerator physicists and technical personnel can meet and exchange
their ideas and new developments.  These PATPAA conferences, conducted 
once in three three years, attract a large participation from across
the entire nation and some from abroad with enthusiasm and 
interest~\cite{PATPAA}.

\medskip

The above Schools are extremely significant since, the 
{\em Accelerator \& Beam Physics and associated technologies} are
{\bf not yet} part of the regular university curriculum in most parts 
of the world!  The learning of such an important interdisciplinary
science is done to a very large extent individually and through the
very few Schools {\em when and where} available~\cite{ICFA-1998}.  We
need to include accelerator \& beam physics in the regular university
curriculum.  In passing it is to be noted that, there is {\bf yet} to
be an {\bf Accelerator \& Beam Physics Association/Society of India}.
The {\bf Japanese Beam Physics Club}~\cite{JBPC} and the
{\bf Particle Accelerator Society of China}~\cite{PASC} provide the
required Forums in their respective countries.  When created, such an
Association/Society will provide the much awaited 
Forum~\cite{ICFA-1998}, strengthening the accelerator \& beam physics
community nation-wide.  This has been the case in various other areas
of Physics, for a very long time.  Then, why should accelerator \&
beam physics continue to make an exception?

\medskip

\section{The Relocated Synchrotrons}

{\bf Siam Photon Source}:
Recently the Japanese donated a $1.0 GeV$ synchrotron to
Thailand~\cite{PT-August-1999}.  Thus, Asia-Pacific region became the
birth-place for the {\em Era of the Relocated Synchrotrons}.  Located
$250 km$ northeast of Bangkok in the city of Nakhon Ratchasima, the
``Siam Photon Source'' is Thailand's first synchrotron light facility
and is intended to serve scientists throughout Southeast Asia.  The
original synchrotron light source, called SORTEC, was located in
Tsukuba Science City, near KEK, Japan's High Energy Accelerator 
Research Organization.  Thailand's Ministry of Science, Technology,
and Environment got the machine {\em gratis} and is investing about 
US~\$~15 million to move and upgrade it.  This includes the doubling 
of the circumference to $81 m$ and tailoring the machine to produce
narrow bright beams of soft X-rays and ultraviolet radiation. 
Scientists from the KEK have helped in the redesign and are
training the scientists from Thailand to operate their new facility.  
Professor Tokehiko Ishii, the retired Director of the Synchrotron
Radiation Laboratory at University of Tokyo is the key figure in
orchestrating the donation.  He is also overseeing the technical and
scientific aspects of the transfer and upgrading of the synchrotron. 
The plan is to use the Siam Photon Source for physics and chemistry
research, with some industrial research in semiconductors, medicine,
pharmaceuticals, and agriculture.  Siam Photon Source is scheduled to
go on-line in 2001~\cite{Siam-Website}. 

\medskip

{\bf SESAME}:
Jordan is the {\em first} country from the Middle East to join the
elite group of countries possessing a synchrotron light 
source~\cite{Nature-July-2000}.  This became possible as Germany 
decided to generously gift the BESSY-I, a $800 MeV$ synchrotron, fully
functioning since 1982 in Berlin, to the region of Middle East.  BESSY
stands for {\em Berliner Elektronen-Spiecherring f\"{u}r
Synchrotronstrahlung}.  BESSY-I is to be replaced by the more powerful
BESSY-II, a $1.9 GeV$ synchrotron located in another part of Berlin.  
Germans are well-known for their environmentally responsible attitude 
towards reusing and recycling, and now they have very successfully
extended that attitude to the large-scale research facilities!  The 
idea of donating the BESSY-I Synchrotron came from Herman Winick of 
the Stanford Linear Accelerator Center (SLAC) in California, a member
of the Machine Advisory Committee of BESSY-II, and the fellow
committee member Gustav-Adolf Voss, a former director of Deutsches 
Elektronen-Synchrotron (DESY) in Hamburg, Germany.  The Project is
known by the acronym {\bf SESAME} (Synchrotron-light for Experimental 
Science and Applications in the Middle East)~\cite{SESAME-Website}. 
The SESAME Project reached a major milestone with the selection of a
site in Jordan at a Meeting of the SESAME Interim Council in Amman,
Jordan during 21-22 June 2000.  SESAME will be the upgraded 
reincarnation of BESSY-I.  The upcoming joint synchrotron radiation
facility, which would be the first regional centre for cooperation in
basic research in the Middle East will also serve as a seed for an
International Centre built around the facility.  SESAME will be 
located at the Al-Balqa' Applied University in Al-Salt and will be 
open to scientists from any country in the region or elsewhere.  
Because of this openness, organizers see its potential as not only a 
world-class research centre, but also as a politically important 
example of scientific cooperation in the region.  Such a centre has
been long overdue and it shall be the first of its kind in the region. 
The Centre will be operated and supported by its eleven member 
countries (Armenia, Cyprus, Egypt, Greece, Iran, Israel, Jordan, 
Morocco, Oman, Palestine and Turkey) with support from countries 
including, France, Germany, Italy, Japan, Russia, Sweden, Switzerland 
and USA.  Other countries which have expressed an interest to join 
this new fount of science and medium of international cooperation 
include, Bahrain, Tunisia and Yemen.  It is hoped that the new centre 
will be able to mirror the CERN in stimulating regional research
collaboration.  Very much like CERN, SESAME is under the very valuable 
political umbrella of UNESCO and is expected to promote science and 
foster international cooperation~\cite{CERN-March-2000}.  A 
broad-spectrum of planned research programmes include, structural
molecular biology, molecular environmental science, surface and
interface science, micro-electromechanical devices, X-ray imaging,
archaeological microanalysis, materials characterization, and medical
applications.  A detailed account of events leading to the SESAME are
available in~\cite{Khan-ICFA,Khan-AAPPS}.

\medskip

{\bf DELSY}: 
A Dutch accelerator and storage ring used for nuclear physics is being
moved to Dubna, to add to Russia's Synchrotron
capability~\cite{Science-June-1999}.  The original facility, was 
located at the Institute of Nuclear Physics and High Energy Physics
(NIKHEF) in Amsterdam, The Netherlands.  This shall be the $1.2 GeV$
``Dubna Synchrotron Radiation Source (DELSY)'', located at the
Joint Institute of Nuclear Research ({\bf JINR}) in Dubna.

\medskip

\section{Asian Committee for Future Accelerators}
The pace of development and the increasing role of accelerator-based
science in Asia has led to the establishment of a forum, the
{\em Asian Committee for Future Accelerators} ({\bf ACFA}), to
discuss and implement plans for further promoting collaborative
accelerator-based science in Asia.  The primary purpose of ACFA is to
strengthen regional collaboration in accelerator-based science, to
encourage future projects in Asia, and to make recommendations to
governments.  This organization was formed in 1996 and its members now
include, China, India, Indonesia, Japan, Korea, Malaysia, Pakistan,
Singapore, Taiwan, Thailand and Vietnam.  The continent of Australia is
also a member~\cite{ACFA-Website}.  The First Asian Particle Accelerator
Conference (APAC-1998) was held at the Japanese High Energy Accelerator
Organization ({\bf KEK}) under the auspices of the ACFA, stressing the
importance of regional collaboration among Asian regions in the field
of accelerator science and technology as well as accelerator-based
science.  The ``Next APAC'' is scheduled to be held during 17-21
September 2001 at Beijing, China~\cite{APAC-Website}.  Globally
speaking, the International Committee for Future Accelerators
({\bf ICFA})~\cite{ICFA-Website}, would provide an excellent framework
for collaborations and forums.  ACFA closely collaborates with ICFA,
a longstanding organization of the world community.  It is noteworthy 
to see how the ICFA Beam Dynamics Panel has contributed to the
accelerator \& beam physics.  The well-attended and very regularly held
ICFA Beam Dynamics Workshops are one of the proofs of its grand success.

\medskip

\section{Concluding Remarks}
We have briefly discussed the synchrotron radiation and the synchrotron
facilities, with a coverage to accelerator programmes in India. 
Siam, SESAME and DELSY are very unique facilities as they are being
built by relocating the very generously donated synchrotrons.  There
are several countries which are in the process of building their own
synchrotrons.  Armenia is planning to build the $3.2 GeV$
{\bf CANDLE}: {\em Center for the Advancement of Natural Discoveries
using Light Emission}~\cite{CANDLE-Website}.  There is the proposal to
build the $3.0 GeV$ {\bf BOOMERANG}~\cite{BOOMERANG-Website} under 
the {\em Australian Synchrotron Research Programme} 
({\bf ASRP})~\cite{ICFA-Australia}.  Spain has the project for a
$2.5 GeV$ {\em National Synchrotron Laboratory} ({\bf LLS}) at
Barcelona~\cite{LLS-Website}.
The upcoming synchrotron facilities will be able to bridge the gap
in several of the under-represented regions of the 
{\em World Synchrotron Map}~\cite{Synchrotron-Map}.  These, when
built, will immensely benefit the scientific community in the 
concerned regions by enhancing international cooperation and providing
them the latest technological expertise.  Among the upcoming
synchrotrons, SESAME is the most international project and offers
an excellent opportunity for participation and active international
collaboration.  
India could have participated and can still get involved
and play a significant role~\cite{Khan-CS}.



\end{document}